\documentclass[aps,prl,10pt,twocolumn,superscriptaddress,balancelastpage,showpacs,reprint,english]{revtex4-1}
\usepackage[T1]{fontenc}
\usepackage[latin9]{inputenc}
\usepackage{amsmath}
\usepackage{amssymb}
\usepackage{graphicx}
\usepackage{color}
\usepackage{float}

\makeatletter

\newcommand{\qq}{\begin{eqnarray}}
\newcommand{\qqq}{\end{eqnarray}}

\newcommand{\bfr}{\mathbf{r}}

\newcommand{\bfJ}{{\bf J}}
\newcommand{\bfq}{{\bf q}}

\pdfpageheight\paperheight
\pdfpagewidth\paperwidth

\makeatother

\usepackage{babel}
\begin{document}

\title{Strong coupling in conserved surface roughening: A new universality class?
}

\author{Fernando Caballero} 
\affiliation{DAMTP, Centre for Mathematical Sciences, University of Cambridge, Wilberforce Road, Cambridge CB3 0WA, UK}
\email{fmc36@cam.ac.uk}

\author{Cesare Nardini} 
\affiliation{Service de Physique de l'\'Etat Condens\'e, CNRS UMR 3680, CEA-Saclay, 91191 Gif-sur-Yvette, France}
\email{cesare.nardini@gmail.com}

\author{Fr\'ed\'eric van Wijland} 
\affiliation{Laboratoire Mati\`ere et Syst\`emes Complexes, UMR 7057 CNRS/P7, Universit\'e Paris Diderot, 10 rue Alice Domon et L\'eonie  Duquet,  75205 Paris cedex 13,  France}
\email{frederic.van-wijland@univ-paris-diderot.fr}

\author{Michael E. Cates} 
\affiliation{DAMTP, Centre for Mathematical Sciences, University of Cambridge, Wilberforce Road, Cambridge CB3 0WA, UK}

\date{\today}

\begin{abstract}
The Kardar-Parisi-Zhang (KPZ) equation defines the main universality class for nonlinear growth and roughening of surfaces. But under certain conditions, a conserved KPZ equation (cKPZ) is thought to set the universality class instead. This has non-mean-field behavior only in spatial dimension $d<2$. We point out here that cKPZ is incomplete: it omits a symmetry-allowed nonlinear gradient term of the same order as the one retained. Adding this term, we find a parameter regime where the $1$-loop renormalization group flow diverges.
This suggests a phase transition to a new growth phase, possibly ruled by a strong coupling fixed point and thus described by a new universality class, for any $d>1$. In this phase, numerical integration of the model in $d=2$ gives clear evidence of non mean-field behavior.
\end{abstract}


\maketitle

Kinetic roughening phenomena arise when an interface is set into
motion in the presence of fluctuations. 
The earliest theoretical investigations \cite{peters1979radius,plischke1984active,jullien1985scaling} were concerned with the Eden model \cite{eden1958}, originally
proposed to describe the shape of cell colonies, and with the ballistic deposition model \cite{family1985scaling}. 
Kardar, Parisi and Zhang (KPZ)~\cite{KPZ} discovered an important universality class for growing rough interfaces, by 
adding the lowest order nonlinearity to the continuum Edwards-Wilkinson (EW) model in which height fluctuations are driven by non-conserved noise and relax diffusively~\cite{edwards1982surface}.
The KPZ equation inspired many analytic, numerical and experimental studies \cite{Krug1997,CorwinKPZreview,Takeuchi:17Rev}, and continues to surprise researchers ~\cite{canet2010nonperturbative,sasamoto2010one,CorwinKPZreview,kriecherbauer2010pedestrian,derrida2007non,meerson2016large}, not least because of a strong-coupling fixed point not accessible perturbatively~\cite{KPZ}. Several experiments have been performed~\cite{takeuchi2014experimental} to confirm the KPZ universality class and recently gained sufficient statistics to show universal properties beyond  scaling laws~\cite{takeuchi2011growing,wakita1997self,maunuksela1997kinetic}.
Finally, the KPZ equation is the first case where solutions to a non-linear stochastic partial differential equation have been rigorously defined ~\cite{hairer2013KPZ}, using a construction related to the Renormalization Group (RG)~\cite{kupiainen2017renormalization}.

Despite its fame, the KPZ equation does not describe all isotropically roughening surfaces; various other universality classes have been identified~\cite{Krug1997,barabasi1995fractal}. In particular, it is agreed that in some cases such as vapor deposition and idealized molecular beam epitaxy~\cite{krim1995experimental}, surface roughening should be described by conservative dynamics (rearrangements dominate any incoming flux), with no leading-order correlation between hopping direction and local slope. These considerations eliminate the EW linear diffusive flux and make the geometric nonlinearity addressed by KPZ not allowed~\cite{Krug1997}. What remains is a conserved version of KPZ equation (cKPZ)~\cite{Sun1989,wolf1990growth,sarma1991new}, which has been widely studied for nearly three decades~\cite{Krug1997,constantin2004persistence,Janssen1996,Racz1991,yook1998correct}:
\qq\label{eq:ckpz}
\dot{\phi}=-\nabla\cdot \bfJ_{\lambda} +{\eta}\quad ; \quad \bfJ_{\lambda} =\nabla\left\{ \kappa\nabla^2\phi+\lambda|\nabla\phi|^2\right\}.
\qqq
Here  $\phi({\bf r},t)$ is the height of the surface above point ${\bf r}$ in a d-dimensional plane, $\bfJ_{\lambda}$ is the deterministic current and
 $\eta$ is a Gaussian conservative noise with variance $\langle \eta(\bfr, t)\eta(\bfr', t')\rangle =-2D\nabla^2\delta^{d}(\bfr-\bfr')\delta(t-t')$. In the linear limit, $\lambda=0$, (\ref{eq:ckpz}) reduces to the Mullins equation for curvature driven growth (a conserved counterpart of EW)~\cite{Mullins} whose large-scale behavior is controlled by two exponents, $\chi=(2-d)/2$ and $z=4$, with spatial and temporal correlators obeying $\langle \phi(\bfr, t)\phi(\bfr',t)\rangle\sim |\bfr-\bfr'|^{2\chi}$ and $\langle \phi(\bfr, t)\phi(\bfr,t')\rangle\sim |t-t'|^{2\chi/z}$. The nonlinear term $\lambda|\nabla\phi|^2$ can be interpreted microscopically as a nonequilibrium correction to the chemical potential, causing jump rates to depend on local steepness at the point of take-off as well as on curvature~\cite{Krug1997}.

\begin{figure}
	\includegraphics[scale=0.26]{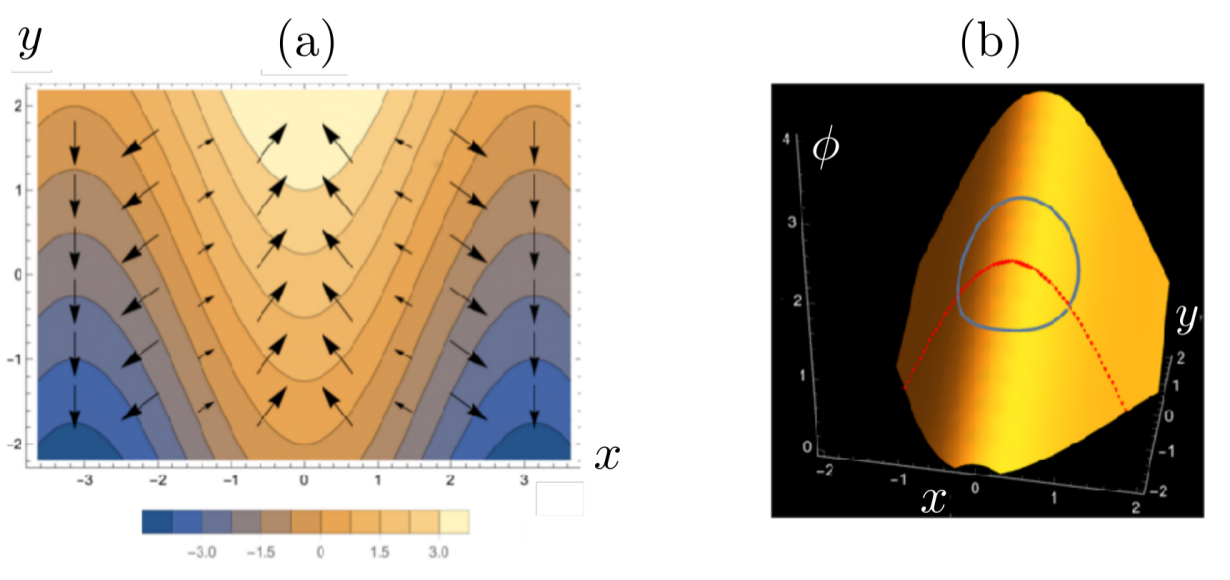}
	\caption{\label{fig:curly}
	(a) Contour plot of the surface $\phi(x,y)=2\cos x +y$ and the current $\bfJ_{\zeta}= -\zeta (\nabla^2\phi) \nabla\phi$ (vectors).  $\bfJ_{\zeta}$ resembles a shear flow and thus has non-zero curl. In (b), the blue line is the locus of points $\cal C$ on the surface equidistant from the origin in the natural metric of the surface. The red dotted line is the intersection between the vertical plane $y=0$ and the surface: because it does not cut $\cal C$ in half, it induces $\bfJ_{\zeta}\neq 0$ if a particle in $(0,0)$ jumps with equal probability to any site on $\cal C$.}
\end{figure}

The properties of the cKPZ universality class are well known \cite{Krug1997}: the upper critical dimension is $2$, above which the RG flow leads to the Gaussian fixed point of the Mullins equation, where $\chi<0$ implies smooth growth. Only for $d<2$ is the nonlinearity relevant; a nontrivial fixed point then emerges perturbatively (see Fig~\ref{fig:lambdazetaflow}(a)). The $1$-loop RG calculation \cite{Sun1989,Janssen1996} shows that, at leading order in $\epsilon=2-d$,  the critical exponents are $z=4-\epsilon/3$ and $\chi=\epsilon/3$; the surface is now rough ($\chi>0$). 
Such predictions turn out to be very accurate when tested against numerical integration of cKPZ~\cite{Krug1997}.

In this Letter we argue that cKPZ is not the most general description of conservative roughening without leading-order slope bias, and that a potentially important universality class may have been overlooked by assuming so. We show this by establishing the importance in $d>1$ of a second, geometrically motivated nonlinear term, also of leading order, whose presence fundamentally changes the structure of the $1$-loop RG flow, creating a separatrix beyond which the flow runs away to infinity. This might lead to three conclusions: $i)$ the runaway is an unphysical feature of the $1$-loop RG flow, cured at higher orders; $ii)$ the separatrix in the RG flow marks a phase transition towards a new phase, where scale invariance is lost; $iii)$ scale invariance is present in this new phase, but its properties are dictated by a strong coupling fixed point. Observe that $iii)$ closely resembles what is found for KPZ whose strong-coupling regime is long-established. We finally perform numerical simulations in the most physically relevant case of $d=2$ and show evidence that the separatrix is not just an artefact of the $1$-loop RG flow. 

For non-conserved dynamics, the KPZ nonlinearity stands alone at leading order after imposing all applicable symmetries. 
For conserved dynamics, however, the cKPZ choice $\bfJ_\lambda$ of deterministic current in \eqref{eq:ckpz} is not the only one possible. All symmetries consistent with $\bfJ_\lambda$ also admit, at the same order ($\nabla^3,\phi^2$), a second term:
\begin{equation}
\bfJ_{\zeta}= -\zeta (\nabla^2\phi) \nabla\phi. \label{eq:zeta}
\end{equation}
A feature of  $\bfJ_{\zeta}$ is that it has nonzero curl, see Fig.~\ref{fig:curly}. The rotational part of any current has no effect on $\dot\phi$ (because $\nabla\cdot\nabla \times \bfJ \equiv 0$), but $\zeta\neq 0$ also means that the irrotational  deterministic current $\bfJ_{\rm irr}$ cannot be expanded in gradients. Put differently, writing $\bfJ_{\rm irr} = \nabla\psi[\phi]$, $\psi$ is {\em not} of the cKPZ form because the Helmholtz decomposition of a vector field does not commute with its gradient expansion.

The $\zeta$ term can be explained by considering more carefully the `blind jumping' dynamics often used to motivate cKPZ~\cite{Krug1997}. Specifically, we suppose jumping particles to move a {\em small fixed geodesic distance} along the surface in a random direction. To visualise the resulting physics, 
consider curving a flat sheet of paper into a sinusoidally corrugated surface $a \cos kx$ and then applying a shear deformation in the $(y,\phi)$ plane to give $\phi(x,y) = a \cos kx + by$.
This resembles a sloping roof with alternating ridges and grooves (Fig.\ref{fig:curly}).
The locus of points of constant geodesic distance from some departure point (with $y = y_0$) is as shown in Fig 1(b).
 We now ask the fraction $f$ of landing sites (i.e., of points on the folded circle) that have $y>y_0$. It can be confirmed  that $f>1/2$ for a point on a ridge ($\phi_{xx}<0$), but $f<1/2$ for a point in a groove ($\phi_{xx}>0$). The resulting bias towards a positive or negative $y$ increment is bilinear in tilt and curvature, vanishing by symmetry when either $k$ or $b$ is zero. It follows that the local deterministic flux in the $y$ direction contains a term $\sim \phi_{xx}\phi_y$ which is not captured by $\lambda$ but demands existence of the $\zeta$ term. This argument generalizes directly to any case where the `landing rate' depends on geodesic distance only. 

We have thus confirmed that the $\zeta$ term is physical, although of course our `blind geodesic jumping' is not the only possible choice of dynamics. With this choice, the $\zeta$ nonlinearity is purely geometric, arising from the transformation from normal to vertical coordinates. Yet the same is true for the KPZ nonlinearity~\cite{KPZ,Krug1997}. 

In summary, for $d>1$, cKPZ is an incomplete model. Its generalization, which we call cKPZ+, reads:
\qq\label{eq:ckpz+}
\dot{\phi}=-\nabla^2 \left\{ \kappa\nabla^2\phi+\lambda|\nabla\phi|^2\right\}
-\nabla\cdot\bfJ_{\zeta}
+{\eta}\;.
\qqq
We have seen no previous work on \eqref{eq:ckpz+}  in the literature. 
Standard dimensional analysis \cite{Tauber2014} shows both $\lambda$ and $\zeta$ to be perturbatively irrelevant for $d>2$, but this does not preclude important differences in critical behavior between cKPZ and cKPZ+ in $d>1$.  We now present strong evidence for this outcome, first by analysing the RG flow perturbatively close to the Gaussian fixed point, where we may hold $\kappa,D$ constant \cite{Tauber2014}, so the RG flow is derived in terms of the reduced couplings
$\bar\lambda^2=\lambda^2D/ \kappa^{3}$ and 
${\bar\zeta}^2= \zeta^2D / \kappa^{3}$.
Transforming (\ref{eq:ckpz+}) into Fourier space with wavevector $\bfq$ and frequency $\omega$, we have
\begin{align}\label{eq:KPZFourier}
\phi(\hat{\bfq})=\phi_0(\hat{\bfq})+\frac{G_0(q,\omega)}{2}\int_{\hat{\bfq}'}g(\bfq,\bfq')\phi(\hat{\bfq}')\phi(\hat{\bfq}-\hat{\bfq}')
\end{align}
where $\hat{\bfq}=(\omega,\bfq)$, $\phi_0(\hat{\bfq})=G_0(q,\omega)\eta(\hat{\bfq})/q^2$, the  bare propagator is 
$G_0(q,\omega) = q^2/ (-i\omega +\kappa q^4)$ and $\eta$ is Gaussian noise with $\langle\eta(\hat{\bfq})\eta(\hat{\bfq}')\rangle = 2Dq^2(2\pi)^{d+1}\delta^{d+1}(\hat{\bfq}+\hat{\bfq}')$. 

In (\ref{eq:KPZFourier}), the nonlinearities $\lambda$ and $\zeta$ enter via a function $g(\bfq,\bfq')$ that, on symmetrising $\bfq'\leftrightarrow(\bfq-\bfq')$, reads 
\qq\label{eq:vertex-g}
g(\bfq,\bfq') &=& -2\lambda \bfq'\cdot(\bfq-\bfq')+\\
&&\zeta\left[\frac{q'^2\bfq\cdot(\bfq-\bfq')}{q^2}\nonumber
+\frac{|\bfq-\bfq'|^2 \bfq\cdot \bfq'}{q^2}\right]\,.
\qqq
We denote the two-point correlation function of the Mullins equation by $C_0(\hat{\bfq},\hat{\bfq}')=(2\pi)^{d+1} C_0(q,\omega) \delta^{d+1}(\hat{\bfq}+\hat{\bfq}')$ where $C_0(q,\omega) = 2D G_0(q,\omega) G_0(-q,-\omega)/q^2$.

\begin{figure}
	\includegraphics[scale=0.23]{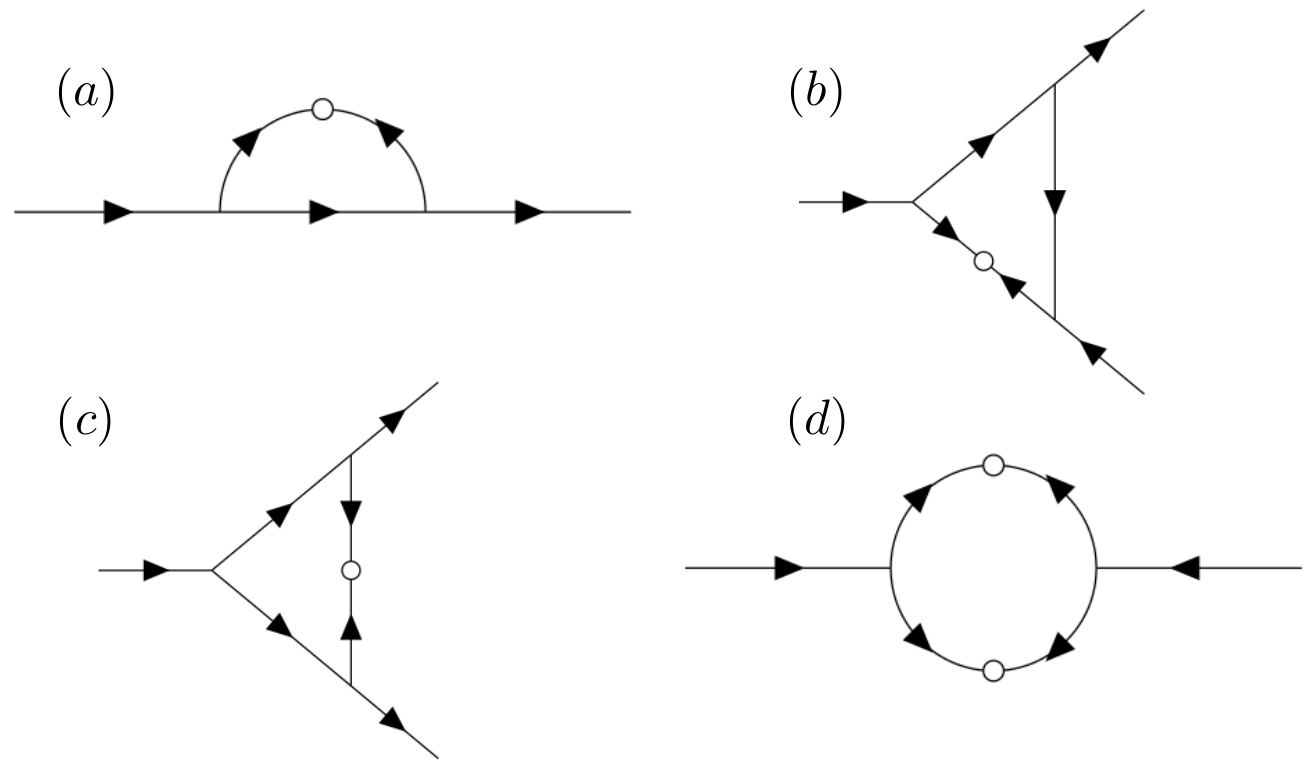}
	\caption{\label{fig:surfacediagrams} The diagrams present at $1$-loop. Those in (b), (c), (d) do not generate any relevant coupling for $\epsilon $ small and close to the Gaussian fixed point.
	\label{fig:diagrams}}
\end{figure}

It is useful to introduce diagrammatic notation, where a line denotes a zeroth-order field $\phi_0$ and the correlation function $C_0(q,\omega)$ is represented as a circle between two incoming lines. The vertex reads $(G_0(q,\omega)/2) \int_{\hat{\bfq}'}g(\bfq,\bfq')\phi(\hat{\bfq}')\phi(\hat{\bfq}-\hat{\bfq}')$,
with $\hat{\bfq}$ the wavevector entering into the vertex.
At $1$-loop, all four diagrams shown in Fig.~\ref{fig:diagrams} might contribute, but a number of simplifications occur. First consider the diagram in Fig.~\ref{fig:diagrams}(d), which could renormalize $D$. Taylor expanding, one finds that the leading contribution is $\mathcal{O}(q^4)$, rendering this irrelevant for small $\epsilon$ and close to the Gaussian fixed point. Next, the two triangular diagrams in Fig. \ref{fig:diagrams}(b) and \ref{fig:diagrams}(c) might renormalize the couplings $\lambda$ and $\zeta$, but explicit computations~\cite{supp}, shows that their contributions exactly cancel out. This can also be shown more directly, by 
 generalising the argument of \cite{Janssen1996}.
 We note that, while $D$ remains un-renormalized at any order in perturbation theory, $\lambda$ and $\zeta$ do get renormalized at higher order. Indeed, this is already known to happen in cKPZ~\cite{Janssen1996}.

We conclude that the diagram in Fig. \ref{fig:diagrams}(a) is the only non-vanishing one to $1$-loop. Its contributions at order $q^0$ and $q$ vanish~\cite{supp}, giving a leading order correction $\mathcal{O}(q^2)$, which renormalizes $\kappa$. Higher terms are irrelevant for small $\epsilon$ and close to the Gaussian fixed point, so we neglect them. The shifted value of $\kappa$ is derived in~\cite{supp} as:
\begin{equation}
\kappa_I=\kappa\left(1-M(\bar\lambda,\bar\zeta,d)\frac{S_d}{(2\pi)^d}\int_{\Lambda/b}^{\Lambda}x^{d-3}dx\right),
\end{equation}
where $S_d=2\pi^{d/2}/\Gamma(d/2)$, 
$(\Lambda/b,\Lambda)$ for $b>1$ is the momentum shell integrated out,
 and
\qq\label{eq:Mfirst}
M(\bar\lambda,\bar\zeta,d) = \frac{1}{8d(2+d)}\left[(2d^2-3d-2)\bar\zeta^2\right.\nonumber\\
\left.+4d(d+2)\bar\lambda\bar\zeta-4(d+2)\bar\lambda^2\right]\,.
\qqq
Since the integrating does not produce new relevant couplings at $1$-loop, we are justified in excluding all higher terms from (\ref{eq:KPZFourier}) and indeed from the cKPZ+ equation \eqref{eq:ckpz+}. 

The last step to obtain the RG flow is rescaling back to the original cut off $\Lambda$  and reabsorbing all rescalings into the couplings. To do so, 
one must introduce scaling exponents for time and the field, such that when $q$ is rescaled as $q\to b q$, then $\omega \to b^{z}\omega$ and $\phi \to b^{-\chi}\phi$. The critical exponents $z$ and $\chi$ are fixed by imposing stationarity of the RG flow at its fixed points. 
This gives the transformation between original and rescaled couplings.

Taking the infinitesimal limit $b=1+db$, we find the RG flow as \qq\label{eq:RG-flow-int}
&\frac{d\kappa}{db}=\kappa\left(z-4-M(\bar\lambda,\bar\zeta,d)\frac{S_d}{(2\pi)^d}\right), \label{eq:KPZ+flow-k}\label{eq:RG-flow-int-1}\\
&\frac{dD}{db}=D(z-2-d-2\chi), \label{eq:KPZ+flow-D}\label{eq:RG-flow-int-2}\\
&\frac{d(\lambda,\zeta)}{db}=(\lambda,\zeta)(z+\chi-4).\label{eq:RG-flow-int-3}
\qqq
Consistent with proximity to the Gaussian fixed point, we impose $d\kappa/db=dD/db=0$ and use (\ref{eq:RG-flow-int-1}-\ref{eq:RG-flow-int-3}) to obtain
\begin{equation}\label{eq:surfflow}
\frac{d(\bar{\lambda},\bar{\zeta})}{db}=(\bar\lambda,\bar\zeta)\left(\frac{2-d}{2}+\frac{3}{2}M(\bar\lambda,\bar\zeta,d)\frac{S_d}{(2\pi)^d}\right).
\end{equation}
This, the central technical result of this Letter, is the $1$-loop RG flow of the cKPZ+ equation \eqref{eq:ckpz+}. 

\begin{figure}
	\includegraphics[scale=0.22]{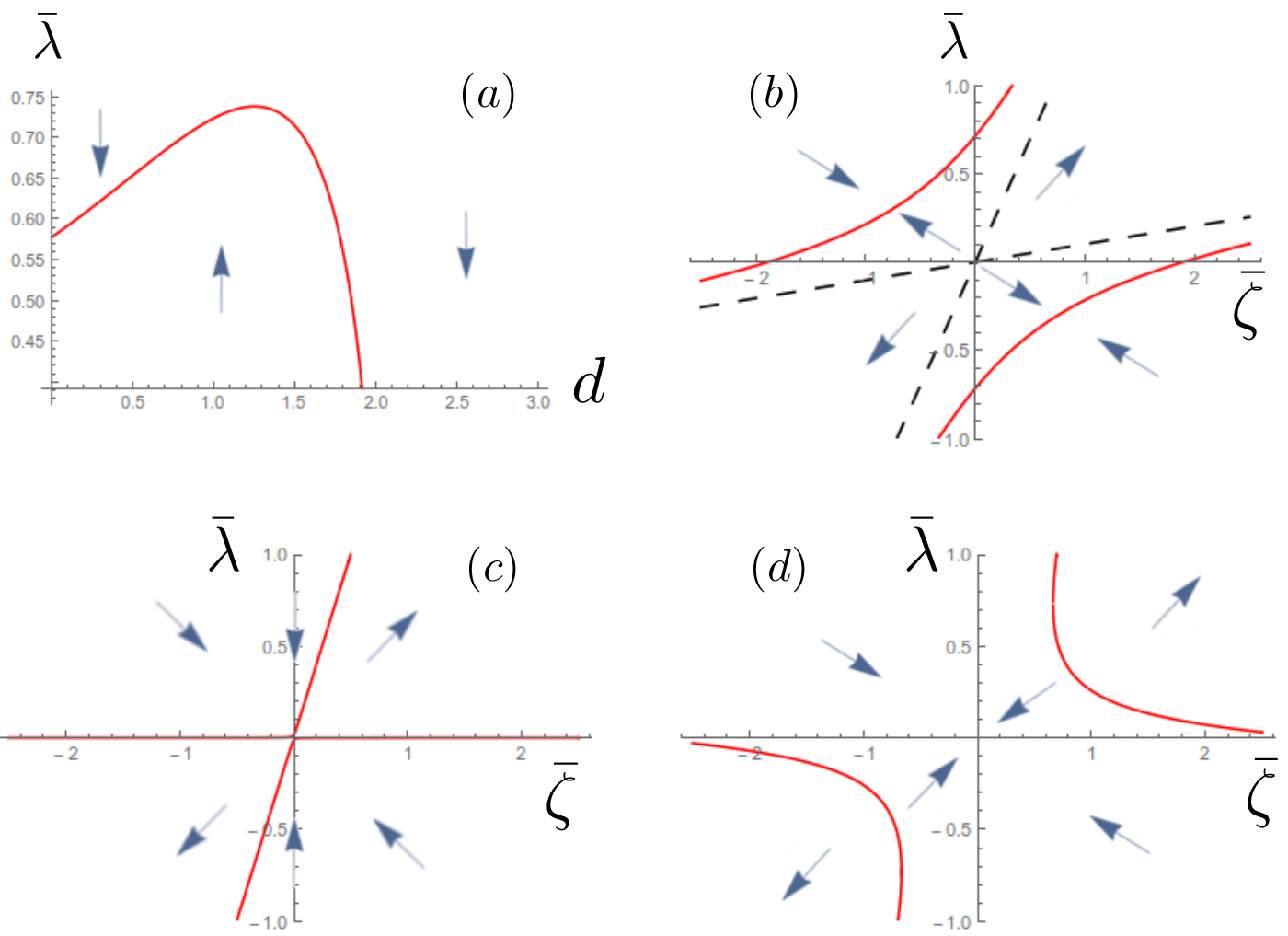}
		\caption{\label{fig:lambdazetaflow}$1$-loop RG flow of the (a) cKPZ model as a function of space dimension $d$ and of the cKPZ+ model for (b) $1<d<2$,  (c) $d=2$,  (d) $d>2$. (cKPZ and cKPZ+ are the same model in $d=1$.)
		The red lines are the fixed points of the RG flow given by solutions of (\ref{eq:KPZ+fixed-conics}) and the dashed lines their asymptotes; the origin is the Gaussian fixed point. In (b),(c),(d), the RG flow is radial and its direction is given by the arrows in the plots.}
\end{figure}

It is now straightforward to obtain the fixed points of the RG flow and their critical exponents setting $d\kappa/db=dD/db=0$ and using (\ref{eq:KPZ+flow-k},\ref{eq:KPZ+flow-D}). First of all, we have the Gaussian fixed point $\lambda=\zeta=0$, whose exponents $z$ and $\chi$ remain those of the Mullins equation mentioned above. 
In the plane of reduced couplings $(\bar\lambda,\bar\zeta)$ we find additional lines of fixed points on the conics defined by 
\qq\label{eq:KPZ+fixed-conics}
S_dM(\bar\lambda,\bar\zeta,d)=(2\pi)^d(d-2)/3\,.
\qqq
Here the critical exponents are those of cKPZ: $z=4-\frac{\epsilon}{3}+\mathcal{O}(\epsilon^2)$ and
$\chi=\frac{\epsilon}{3}+\mathcal{O}(\epsilon^2)$.
The stability of the RG fixed points is shown Fig. \ref{fig:lambdazetaflow}. For $d>2$ the Gaussian point is locally stable (Fig.~\ref{fig:lambdazetaflow}d). However, the two new lines of fixed points are unstable and the basin of attraction of the Gaussian fixed point shrinks when approaching $d\to2^+$. In $d< 2$, the Gaussian fixed point becomes unstable while the lines of fixed points defined by (\ref{eq:KPZ+fixed-conics}) are stable (Fig.~\ref{fig:lambdazetaflow}a,b). Nonetheless, in $1<d\le 2$ the latter are not globally attractive because there are sectors of the reduced couplings plane where the RG flow runs away to infinity. These sectors exclude the  pure cKPZ case ($\zeta =0$, vertical axis) so that the runaway is a specific feature of cKPZ+. Similar remarks apply in $d>2$ where the unstable fixed lines are separatrices between the Gaussian fixed point and a runaway to infinity.

The scenario just reported resembles that of KPZ  at $2$-loops \cite{frey1994two,Tauber2014}. There, the Gaussian fixed point is stable for $d>2$ and unstable for $d<2$. A non-trivial fixed point is again present which is stable for $d<2$ but unstable for $d>2$, where the Gaussian fixed point has a finite basin of attraction, beyond which the flow runs away. In KPZ, this scenario signifies the emergence of a nonperturbative, strong-coupling fixed point~\cite{KPZ,canet2010nonperturbative,frey1994two}, whose existence and properties are by now well established. The two main differences with respect to KPZ are: $i)$ in KPZ, the coupling constant at the non-Gaussian fixed point diverges in the limit  $d\to 2^-$~\cite{Tauber2014} and $ii)$ in KPZ, for $d<2$, the non-trivial fixed point is fully attractive. 
Despite these differences, it is natural to conclude that the runaway to infinity 
signifies the presence of a strong-coupling fixed point, with a distinct universality class, also for cKPZ+ in $d>1$. However, as anticipated in the introduction, two other scenarios are possible: the runaway to infinity might be just an artefact of the $1$-loop computation or the separatrix in the RG flow could signal a phase transition to a different growth phase without scale invariance.

 \begin{figure}
	\includegraphics[scale=0.65]{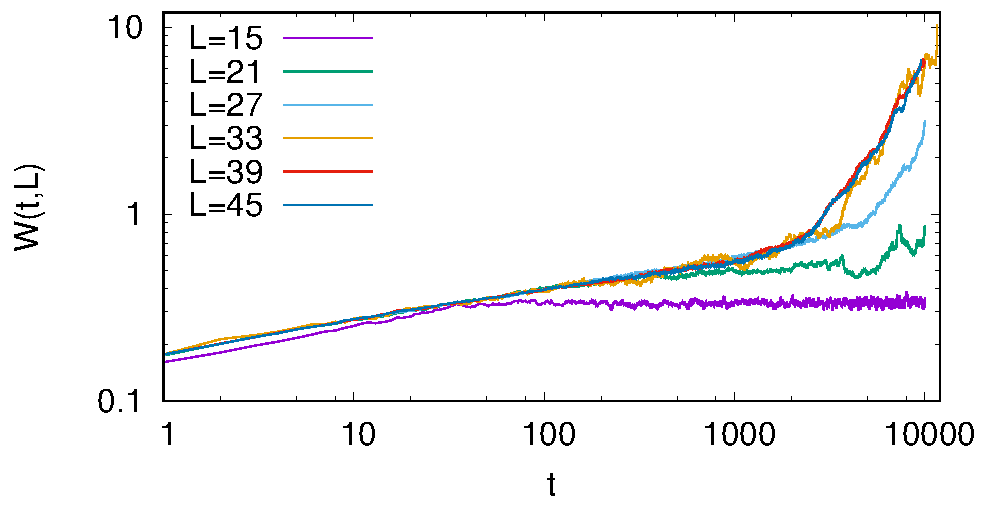}	
	\caption{\label{fig:diverging}  Growth of the width of the interface $W(L,t)$ with time for different system sizes $L$ of the cKPZ+ equation in $d=2$. The parameters $\lambda=1/2$ and $\zeta=1$ were chosen to lie in the region where the RG flow diverges. Each line is an average over several noise realizations (from $1600$ for $L=15$ to $80$ for $L=45$). The value of $k_6=0.2$. The difference between the present case and the one where the RG flow converges towards the Gaussian fixed point is apparent. Here, at late times (decreasing with $k_6$ increasing~\cite{supp}), the width grows faster than logarithmically. The growth law seems independent from system size.
}
\end{figure}

In order to rule out that the runaway of the RG flow is an artefact of the $1$-loop computation, we performed numerical simulations of (\ref{eq:ckpz+}) in $d=2$, the physically most relevant case. We used a  pseudo-spectral code with $2/3$ dealiasing procedure and Heun scheme~\cite{mannella2000gentle} for the time integration. In all simulations, we set $D=\kappa=1$ and all the results shown are obtained starting from a flat initial condition $\phi=0$, but we checked that no difference is obtained when starting from a random initial condition. We checked the stability of our results upon varying the time-step in the window $(10^{-4}, 5\times 10^{-3})$. The system-sizes used are $L\times L$, with $L$ varying between $15$ and $45$. 
As standard in the study of roughening surfaces, we report below results on the width  of the interface $W(L,t)\equiv (1/L^2)\int_{\bfr } \langle \phi^2(\bfr,t) \rangle $. For fixed $L$, we studied the growth of $W$ with time $t$ and the large-time saturated width.

Within the basin of attraction of the Gaussian fixed point, the code proved numerically stable, allowing us to reproduce the expected critical behavior~\cite{supp}: $W(t,L)\sim \log t$ and $z=4$. Moreover, simulations in $d=1$ gave exponents agreeing with the known cKPZ values (not shown). In contrast, for parameters where the RG flow diverges to infinity, in order to obtain numerically stable results, we had to add a higher-order regularizer in the form of $k_6\nabla^6\phi$ in (\ref{eq:ckpz+}). This is irrelevant close to the Gaussian fixed point and does not affect the RG flow there. In Fig.~\ref{fig:diverging}, we report $W(L,t)$ as a function of time for different system sizes. The behavior differs strongly from the mean-field one: after an initial transient, shown in~\cite{supp} to depend on $k_6$, $W$ grows much faster than logarithmically. At large enough system sizes, a seemingly size-independent algebraic growth law emerges, although larger $L$ values would be needed to confirm this. Fig. \ref{fig:diverging} is obtained by averaging over many noise realizations (from $1600$ for $L=15$ to $400$ for $L=45$). We report in ~\cite{supp} the behavior of $W$ for a few individual ones, showing a strong increase in the variance of $W$ with time. This seems to be associated with the late stage algebraic growth regime.

Our simulations give clear evidence that the runaway to infinity is not an artefact of the $1$-loop RG flow. We leave open the question of whether cKPZ+, within certain parameter regimes, has a new universality class or, instead, scale invariance is lost there. In the latter case, the properties of this phase might be linked to a `mounding' phase, seen in conserved roughening surfaces when starting from sufficiently steep initial conditions~\cite{chakrabarti2004mound}.

In summary, we have argued that the cKPZ equation (\ref{eq:ckpz}), thought to govern conserved, slope-unbiased roughening dynamics, is incomplete. We introduced a new model, cKPZ+ (\ref{eq:ckpz+}), with a complete set of leading-order nonlinearities. In $d=1$, cKPZ and cKPZ+ coincide, but they differ in any $d>1$. Surprisingly, the RG analysis of cKPZ+ at $1$-loop suggests the presence of a non mean-field growth phase in any dimension $d>1$, which might be due to a new universality class or a loss of scale invariance. Indeed, our numerical analysis clearly indicates that the runaway of the $1$-loop RG flow signifies new physics at strong coupling.

\begin{acknowledgments}
{\em Acknowledgements.}
FC is funded by EPSRC DTP IDS studentship, project number $1781654$.
CN acknowledges the hospitality provided by DAMTP, University of Cambridge while part of this work was being done.
CN acknowledges the support of an Aide Investissements d'Avenir du LabEx PALM (ANR-10-LABX-0039-PALM). Work funded in part by the European Resarch Council under the EU's Horizon 2020 Programme, grant number 760769. MEC is funded by the Royal Society.
\end{acknowledgments}

\bibliographystyle{apsrev4-1}
\bibliography{biblio}

\newpage
\begin{center}
\textbf{\large Supplemental Information: Strong coupling in conserved surface roughening: A new universality class?}
\end{center}
\section{renormalization of $\kappa$}\label{app-ren-kappa}
We consider here the diagram of Fig.2(a) in the main text and, in particular, the renormalization to $\kappa$ which arises from it. 

Explicitly, the diagram reads 
\qq\label{eq:app:loop-1}
\frac{G_0(q,\omega) }{4}
&&\int_{\hat{\bfq}'}\frac{1}{(2\pi)^{d+1}}
g(\bfq,\bfq')g(\bfq-\bfq',\bfq)\nonumber\\
&&G_0(|\bfq-\bfq'|,\omega-\omega')C_0(\hat{\bfq}')
\phi(\hat{\bfq})
\qqq
We compute now the loop integral of (\ref{eq:app:loop-1}). We observe that we can take the limit $\omega\to 0$ since we are looking at the system in the hydrodynamic limit and thus Taylor expand contributions for small $q$ and $\omega$. The integral over the time frequency $\omega'$ can be readily done via contour integration; we get
\begin{equation}
\int_{-\infty}^{\infty}\frac{d\omega'}{2\pi}G_0(|\bfq-\bfq'|,-\omega')C_0(\hat{\bfq}')=\frac{D}{\kappa^2}\frac{|\bfq-\bfq'|^2}{q'^2\left[|\bfq-\bfq'|^4+q'^4\right]}.
\end{equation}
The loop integral in (\ref{eq:app:loop-1}) thus becomes
\qq
	&&\frac{D}{4\kappa^2(2\pi)^d}\int_{\bfq'}\nonumber\\
	&&\left[-2\lambda \bfq'\cdot(\bfq-\bfq')+\zeta\frac{q'^2  \bfq\cdot(\bfq-\bfq')}{q^2}+\zeta\frac{|\bfq-\bfq'|^2 \bfq\cdot \bfq'}{q^2}\right]\nonumber\\
	&&\left[2\lambda \bfq\cdot \bfq'+\zeta\frac{q'^2\bfq\cdot (\bfq-\bfq')}{|\bfq-\bfq'|^2}+\zeta\frac{q^2(-\bfq')\cdot(\bfq-\bfq')}{|\bfq-\bfq'|^2}\right]\nonumber\\
	&&\frac{|\bfq-\bfq'|^2}{q'^2\left[|\bfq-\bfq'|^4+q'^4\right]},\label{eq:ugly}
\qqq
We must now expand this as a Taylor series in $q$. As mentioned in the main text, the first non vanishing order in (\ref{eq:ugly}) is $q^2$, which renormalizes $\kappa$. Indeed, expanding the first parenthesis $\left[\cdot\right]$ in (\ref{eq:ugly}), no contribution that diverges for small $q$ is found. Moreover, the second parenthesis vanishes for $q=0$. A zeroth order term is thus ruled out. In addition, linear terms in $q$ cannot be obtained, as follows from the fact that $\int_{\bfq'} (\bfq \cdot\bfq')f(q')=0$ for any function $f$. The first contribution coming from (\ref{eq:ugly}) is thus at order $q^2$.
Higher terms would be strongly irrelevant close for $d=2+\epsilon$ with $\epsilon$ small, so they can be neglected.  

We now explicitly compute the $q^2$ contribution of (\ref{eq:ugly}). The angular intergrals can be done in any dimension $d$ using the following trick: we write $\bfq\cdot \bfq'=q q'\cos\theta$, so that the following equalities follow 
\qq\label{eq:app:sphere}
	\int_{\text{Sphere}}\cos^2\theta &=& \frac{S_d}{d}\nonumber\\
	\int_{\text{Sphere}}\cos^4\theta &=& \frac{3S_d}{d(d+2)}.
\qqq
Injecting now (\ref{eq:app:sphere}) and (\ref{eq:ugly}) in (\ref{eq:app:loop-1}), we obtain a shift in $\kappa$ as in (6) of the main text, where
\qq
	&&M\frac{S_d}{(2\pi)^d}=\frac{S_d}{(2\pi)^d}\frac{D}{8d(d+2)\kappa^2}\nonumber\\
	&&\left[(2d^2-3d-2)\zeta^2+4d(2+d)\lambda\zeta-4(2+d)\lambda^2\right].
\qqq
This expression coincides with (7) given in the main text by reabsorbing the factor $D/\kappa^2$ into the couplings to form the reduced ones $\bar\lambda$ and $\bar\zeta$.

%%%%%%%%%%%%%%%%%%%%%%%%%%%%%%%%%%%%%%%%%%%%%%%%%%%%%%%
%%%%%%%%%%%%%%%%%%%%%%%%%%%%%%%%%%%%%%%%%%%%%%%%%%%%%%% 

\section{No renormalization of the couplings to $1$-loop}\label{app-ren-triangular}
 \begin{figure}
		\includegraphics[scale=0.275]{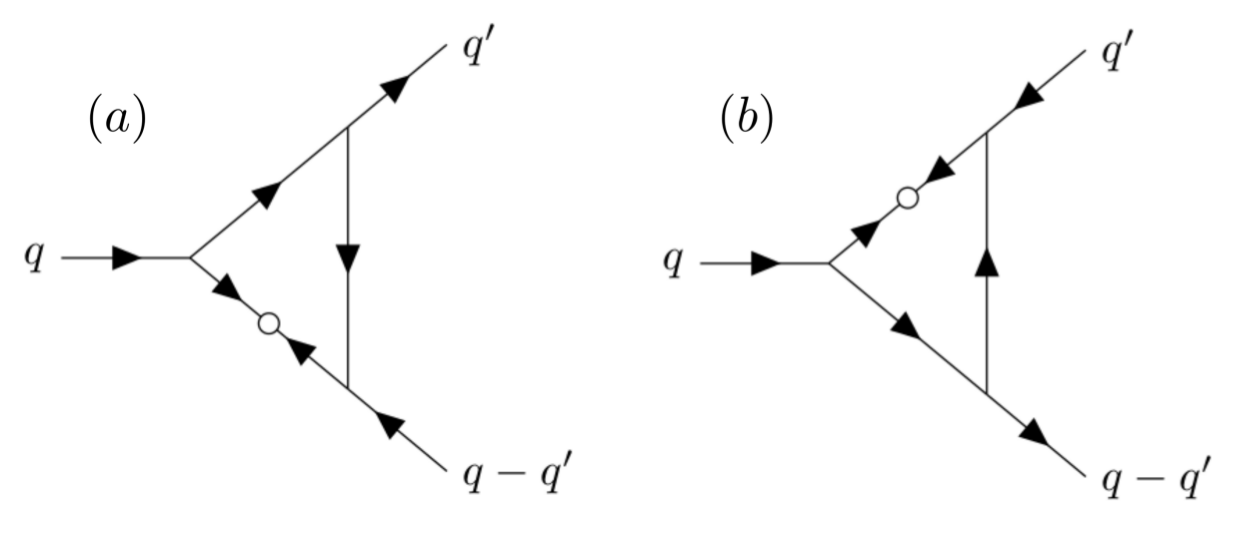}
	\caption{\label{fig:triangular} For simplicity, we split the diagram of Fig.  2(c) of the main text in the two contributions depicted here.
	}
	\end{figure}
As stated there, the contribution of the two triangular diagrams in Fig. 2(b,c) of the main text exactly cancel out. Such a result can be obtained generalising the argument of~\cite{Janssen1996} to the KPZ+ equation. We checked that this result is correct by explicit computation, as sketched below. This means that there is no renormalization of the couplings $\lambda$ and $\zeta$ to $1$-loop.

For simplicity, we will consider the diagrams in Fig. 2(b,c) with an incoming momentum denoted by $\hat{\bfq}$, and two outgoing ones $\hat\bfq'$ and $\hat{\bfq}-\hat{\bfq}'$; we also call $\hat{\bfq}_I$ the internal momentum flowing in the loop. The angles between the vectors $\bfq,\bfq_I$ and $\bfq',\bfq_I$ will be respectively denoted by 
$\theta,\phi$.

The diagram in Fig. 2(c) of the main text gives a contribution 
\qq\label{eq:app:triangular-1}
 D_1=\int_{\hat{\bfq}_I}&&g(\bfq,\bfq'+\bfq_I)g(\bfq'+\bfq_I,\bfq')g(\bfq-\bfq'-\bfq_I,-\bfq_I)\nonumber\\
&&G_0(| \bfq'+\bfq_I |,\omega'+\omega_I) C_0(\hat{\bfq}_I)\nonumber\\
&&G_0(| \bfq-\bfq'-\bfq_I |, \omega-\omega'-\omega_I)\,. 
\qqq
The frequency integral in the above expression, after setting external frequencies to $0$, gives
\qq\label{eq:app:triangular-2}
 \int_{-\infty}^\infty &&\frac{d\omega_I}{2\pi}G_0(| \bfq'+\bfq_I |,\omega'+\omega_I)C_0(\hat{\bfq}_I)\nonumber\\
&&G_0(| \bfq-\bfq'-\bfq_I |, \omega-\omega'-\omega_I)=\nonumber\\
&&=\frac{  a_1 b_2 D \left(a_1^2+b_1^2+2 q_I^4\right)}{\kappa ^3 q_I^2 (a_1^2+b_1^2) \left(a_1^2+q_I^4\right) (b_1^2+q_I^4)}, 
\qqq
with 
\[ a_1=| \bfq_I+\bfq' |^2\qquad b_1=|\bfq-\bfq'-\bfq_I |^2. \]
The contribution $D_1$ is then obtained by substituting (\ref{eq:app:triangular-2}) into (\ref{eq:app:triangular-1}) and expanding up to quadratic order in the magnitude of the external momenta $\bfq,\bfq'$, obtaining:
\qq
D_1=-\frac{1}{2\kappa ^3 q_I^2}\Big[ &&
D q' (\zeta -2\lambda )^2 \cos \phi  (\zeta  \cos 2 \theta -2\lambda ) \nonumber\\
&&(q' \cos \phi - q \cos \theta )
\Big]
\qqq

\begin{figure}[!htb]
	\includegraphics[scale=0.8]{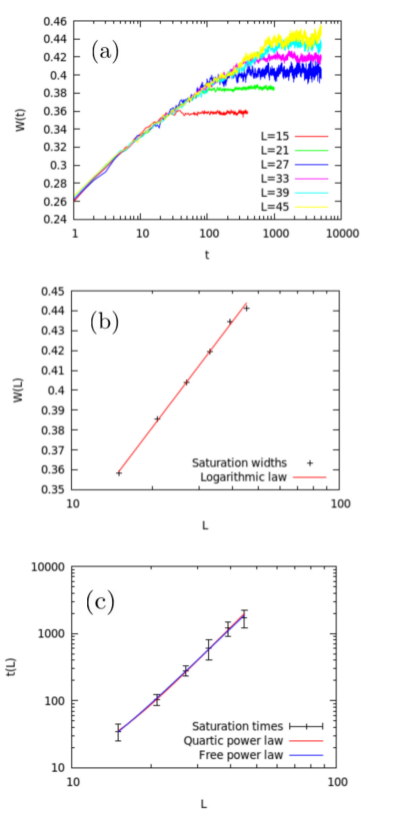}	
	\caption{\label{fig:converging} 
Typical logarithmic growth of a surface for different lattice sizes
$L$ and $\lambda= -1/2 ,\zeta= 2.6$ belonging to the region where the RG flow converges to the Gaussian fixed point. We used $\Delta t=10^{-3}$, $\kappa=D=1$ and averaged the results over many noise realizations ($1600$ for $L=15$ and $80$ for $L=45$). In (a) a logarithmic growth of the width is observed before saturation. In (b) we show that also the saturated value 
$W_{\infty}(L)=\lim_{t\to \infty} W(t,L)$
 as a function of $L$ increase logarithmically (the red line is a logarithmic fit). Both these results are those expected at the Gaussian fixed point in $d=2$. In (c), we report the saturation time $t_s(L)$, which has been estimated from the curves in (a), as a function of the size. $t_s(L)$ is expected to grow as $L^z$ with $z=4$; the best fit gives $z=3.8\pm 0.2$. 
	}
\end{figure}
We now consider the diagram in Fig. 2(b) of the main text. As depicted here in Fig. \ref{fig:triangular}, it has two possible arrangements of momenta inside. The one in Fig. \ref{fig:triangular}(a) gives
\qq\label{eq:SM:tri-0}
D_2=
\int_{\hat{\bfq_I}}&&
g(\bfq,\bfq_I)g(\bfq-\bfq_I,\bfq')g(\bfq-\bfq'-\bfq_I,-\bfq_I)\nonumber\\
&&C_0(\hat{\bfq}_I) G_0(| \bfq-\bfq_I |, \omega-\omega_I)\nonumber\\
&&G_0(| \bfq-\bfq'-\bfq_I |, \omega-\omega'-\omega_I) .
\qqq
Again we set external frequencies to $0$ and integrate over the internal one
\qq\label{eq:SM:tri-2}
\int_{-\infty}^\infty
\frac{d\omega_I}{2\pi}
&&C_0(\hat{\bfq}_I)
G_0(| \bfq-\bfq_I |, \omega-\omega_I)\nonumber\\
&&G_0(| \bfq-\bfq'-\bfq_I |, \omega-\omega'-\omega_I)\nonumber\\
&&= \frac{  a_2 b_2 D}{\kappa ^3 q_I^2 \left(a_2^2+q_I^4\right) \left(b_2^2+q_I^4\right)}, 
\qqq
where
\qq
a_2=| \bfq-\bfq_I |^2\qquad b_2=| \bfq-\bfq'-\bfq_I |^2.
\qqq
Plugging (\ref{eq:SM:tri-2}) into (\ref{eq:SM:tri-0}) and expanding to quadratic order, 
\qq
D_2=\frac{Dq'}{4\kappa^3 q_I^2}
&&\Big[
q' \left(2 \zeta  \cos ^2\theta -\zeta -2\lambda \right) \cos^2\phi\,(\zeta   -2\lambda  )^2\nonumber\\
&&
-q (\zeta -2\lambda )^2 \cos \phi\cos \theta  \left(2 \zeta  \cos ^2\theta -\zeta   -2\lambda   \right)\nonumber
\Big].
\qqq

Analogously, we can compute the diagram in Fig. \ref{fig:triangular}(b). Its explicit computation, very similar to the previous one, is not reported here and gives $D_3=D_2$. It is then straightforward to observe that $D_1+D_2+D_3=0$, meaning that the sum of the diagrams in Fig. 2(b,c) of the main text exactly cancel.

%%%%%%%%%%%%%%%%%%%%%%%%%%%%%%%%%%%%%%%%%%%%%%%%%%%%%%%
%%%%%%%%%%%%%%%%%%%%%%%%%%%%%%%%%%%%%%%%%%%%%%%%%%%%%%% 
\section{Weak Coupling Region}
In this Section we present the numerical results obtained with the pseudo-spectral algorithm in $d=2$ when setting the parameters $\lambda, \zeta$ in the region where the RG flows to the Gaussian fixed point. Here, $k_6=0$ but we checked that the results obtained are independent of it. As shown in Fig.~\ref{fig:converging}, we are able to extract rather accurately the expected behaviour of the Mullins equation. Details are given in the caption of Fig.~\ref{fig:converging}.

 \begin{figure}[!h]
	\includegraphics[scale=0.55]{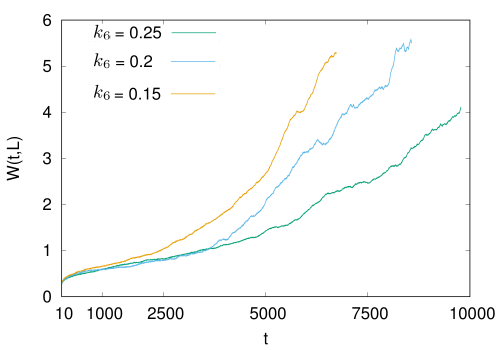}	
	\caption{\label{fig:time-scale-k6} 
Simulations of cKPZ+ in $d=2$ for $\lambda=1/2$ and $\zeta=1$, belonging to the region where the RG flow diverges. 
The three curves where obtained for a system of linear size $L=33$ and three different values of $k_6$. The crossover to the late stage of the growth is shifted to later times when increasing $k_6$. 	}
\end{figure}

%%%%%%%%%%%%%%%%%%%%%%%%%%%%%%%%%%%%%%%%%%%%%%%%%%%%%%%
%%%%%%%%%%%%%%%%%%%%%%%%%%%%%%%%%%%%%%%%%%%%%%%%%%%%%%% 
\section{Runaway Region}\label{app:diverging}
Simulations in the parameter region where the RG flow runs away to infinity were performed using $k_6\neq 0$. In Fig.~\ref{fig:time-scale-k6}, we show that the time-scale for the crossover to the late time growth behavior depends on $k_6$: for smaller $k_6$, the crossover takes place at earlier times. Observe that we stop plotting the curves in Fig.~\ref{fig:time-scale-k6} when at least one realization loses stability. This is the reason for which curves at smaller $k_6$ ends at earlier times.

\begin{figure}[H]
	\includegraphics[scale=0.65]{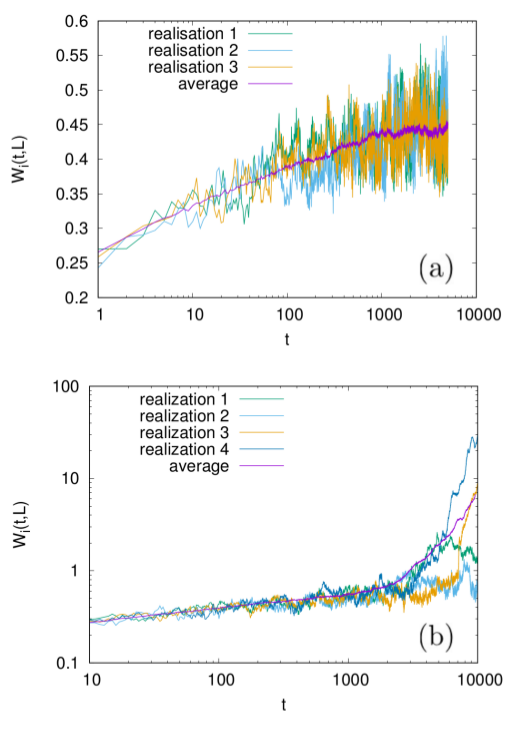}	
	\caption{
{Simulations of cKPZ+ in $d=2$ for (a) $\lambda= -1/2 ,\zeta= 2.6, k_6=0$ (weak coupling) and (b) $\lambda=1/2$ and $\zeta=1, k_6=0.2$ (runaway region). On the $y$-axis, we report the height variance $W_i(t)= (1/L^2)\int_{\bfr }  \phi^2(\bfr,t) $ of individual realizations $i$ and the average over many of them $W(t,L)$.
After the crossover to the late stage supra-logarithmic growth of $W(t,L)$, individual realizations  show a much higher variance in $W_i(t,L)$ than the ones seen at weak coupling. Note the difference in the scale of the $y$-axis in (a) and (b).}
\label{W-individual}	}
\end{figure}

Individual realizations of $W(t,L)$ show a very different behaviour between the weak coupling and the runaway regions. In Fig. \ref{W-individual} we report a few of these, in both regions of parameters. 
At late times, where $W(t,L)$ apparently increases faster than logarithmically with time, we qualitatively observe that individual realizations show a much higher variance than either at earlier times or in the weak coupling regime.

\end{document}